\documentclass[12pt]{article}

\usepackage{color}
\usepackage{a4}
\usepackage{latexsym}
\usepackage{cite}
\usepackage{lineno}

\usepackage{color}
\usepackage{xcolor}

\usepackage{array}
\newcolumntype{T}{>{\ttfamily} c}
\newcolumntype{M}{>{$\displaystyle} c <{$}}

\usepackage{epsfig}
\usepackage{graphicx}
\usepackage{amsmath}
\usepackage{amssymb}
\usepackage{multirow}
\allowdisplaybreaks

\usepackage{pslatex}
\usepackage[latin1]{inputenc}
\usepackage[T1]{fontenc}

\usepackage{color,colordvi}
\def\colour4colour#1{\Blue{#1}}

\newcommand{\gsim}{\raisebox{-0.7mm}{$\:\:\stackrel{>}{{\scriptstyle
 \sim}}\:\: $} }
\newcommand{\lsim}{\raisebox{-0.7mm}{$\:\:\stackrel{<}{{\scriptstyle
 \sim}}\:\: $} }

\def\slash#1{\rlap{\hbox{$\mskip 1 mu /$}}#1} 

\newcommand{\beq}{\begin{equation}}
\newcommand{\eeq}{\end{equation}}
\newcommand{\bea}{\begin{eqnarray}}
\newcommand{\eea}{\end{eqnarray}}
\newcommand{\nn}{\nonumber}
\newcommand{\MSb}{$\overline{\mbox{MS}}$}
\newcommand{\ra}{\rightarrow}

\newcommand{\als}{\alpha_{\rm s}}

\newcommand{\ep}{\varepsilon}

\newcommand{\hspm}{{\hspace{-3mm}}}
\newcommand{\hspn}{{\hspace{-5mm}}}
\newcommand{\hspp}{{\hspace{5mm}}}

\def\as(#1){{\alpha_{\rm s}^{\,#1}}}
\def\ar(#1){{a_{\rm s}^{\,#1}}}

\def\zr#1{{\zeta_{\:\!#1}^{}}}
\def\mus{{\mu^{\,2}}}

\def\B(#1,#2){{\beta_{#1}^{\,#2}}}

\def\col{\color{black}}

\def\nc{{\col{n_c}}}
\def\ncs{{\col{n_{c}^{\,2}}}}
\def\nct{{\col{n_{c}^{\,3}}}}
\def\ncf{{\col{n_{c}^{\,4}}}}

\def\ca{{\col{C^{}_A}}}
\def\cas{{\col{C^{\,2}_A}}}
\def\cat{{\col{C^{\,3}_A}}}

\def\cf{{\col{C^{}_F}}}
\def\cfs{{\col{C^{\, 2}_F}}}
\def\cft{{\col{C^{\, 3}_F}}}
\def\cff{{\col{C^{\, 4}_F}}}
\def\nf{{\col{n^{}_{\! f}}}}
\def\nfz{{\col{n^{\,0}_{\! f}}}}
\def\nfo{{\col{n^{\,1}_{\! f}}}}
\def\nfs{{\col{n^{\,2}_{\! f}}}}
\def\nft{{\col{n^{\,3}_{\! f}}}}

\def\dfRRnR{{\col {d_{\,R}^{\,abcd} d_{\,R}^{\,abcd}\! / n_c} }}
\def\dfRAnR{{\col {d_{\,R}^{\,abcd} d_{\,A}^{\,abcd}\! / n_c} }}

\def\dfRAnr{{\col{ {d_{RA}^{(4)} \over n_c} }}}
\def\dfRRnr{{\col{ {d_{RR}^{(4)}\over n_c} }}}

\def\xm1{{(1 \! - \! x)}}
\def\xp1{{(1 \! + \! x)}}

\def\Lnt(#1){\ln^{\,#1}(1\!-\!x)}
\def\pqq(#1){p_{\rm{qq}}(#1)}
\def\pgq{p_{\rm{gq}}}

\def\bfkl1{{\mbox{bfkl}_1}}
\def\gs{{g_{\rm s}}}
\def\gss{{g_{\rm s}^{2}}}

\def\z#1{{\zeta_{#1}}}

\def\S(#1){{{S}_{#1}}}
\def\Ss(#1,#2){{{S}_{#1,#2}}}
\def\Sss(#1,#2,#3){{{S}_{#1,#2,#3}}}
\def\Ssss(#1,#2,#3,#4){{{S}_{#1,#2,#3,#4}}}
\def\Sssss(#1,#2,#3,#4,#5){{{S}_{#1,#2,#3,#4,#5}}}
\def\Ssssss(#1,#2,#3,#4,#5,#6){{{S}_{#1,#2,#3,#4,#5,#6}}}
\def\Sssssss(#1,#2,#3,#4,#5,#6,#7){{{S}_{#1,#2,#3,#4,#5,#6,#7}}}

\def\Sp(#1,#2){{{S}_{#1}^{\,#2}}}

\def\H(#1){{\rm{H}}_{#1}}
\def\Hh(#1,#2){{\rm{H}}_{#1,#2}}
\def\Hhh(#1,#2,#3){{\rm{H}}_{#1,#2,#3}}
\def\Hhhh(#1,#2,#3,#4){{\rm{H}}_{#1,#2,#3,#4}}
\def\Hhhhh(#1,#2,#3,#4,#5){{\rm{H}}_{#1,#2,#3,#4,#5}}
\def\Hhhhhh(#1,#2,#3,#4,#5,#6){{\rm{H}}_{#1,#2,#3,#4,#5,#6}}

\def\ddelta_{{\delta}}
\def\Dplus(#1){\mathcal{D}_{#1}}
\def\D(#1){{ D_{#1}}}
\def\Dd(#1,#2){{ D_{#1}^{\,#2}}}

\begin{document}
\setlength{\parskip}{0.2cm}
\setlength{\baselineskip}{0.54cm}


\begin{titlepage}
\noindent
ZU-TH 20/24 \hfill April 2024 \\
DESY-24-053 \\
LTH 1367 
\vspace{3mm}
\begin{center}
{\LARGE \bf Four-loop splitting functions in QCD \\[2mm]
-- The quark-to-gluon case --}\\
\vspace{1.5cm}
\large
G.~Falcioni$^{\, a,b}$, F.~Herzog$^{\, c}$, S. Moch$^{\, d}$,
A. Pelloni$^{\, e}$ and A. Vogt$^{\, f}$\\

\vspace{1.0cm}
\normalsize
{\it $^a$Dipartimento di Fisica, Universit\`{a} di Torino,
  Via Pietro Giuria 1, 10125 Torino, Italy}\\
\vspace{1mm}
{\it $^b$Physik-Institut, Universit\"{a}t Z\"{u}rich, 
  Winterthurerstrasse 190, 8057 Z\"{u}rich, Switzerland}\\
\vspace{5mm}
{\it $^c$Higgs Centre for Theoretical Physics, School of Physics and Astronomy\\
  The University of Edinburgh, Edinburgh EH9 3FD, Scotland, UK}\\
\vspace{5mm}
{\it $^d$II.~Institute for Theoretical Physics, Hamburg University\\
\vspace{0.5mm}
Luruper Chaussee 149, D-22761 Hamburg, Germany}\\
\vspace{4mm}
{\it $^e$Institute for Theoretical Physics, 
  ETH Z\"{u}rich, 8093 Z\"{u}rich, Switzerland} \\
\vspace{4mm}
{\it $^f$Department of Mathematical Sciences, University of Liverpool\\
\vspace{0.5mm}
Liverpool L69 3BX, United Kingdom}\\
\vspace{1.4cm}
{\large \bf Abstract}
\vspace{-0.2cm}
\end{center}
We present the even-$N$ moments $N\le 20$ of the fourth-order (N$^3$LO) 
contribution $P_{\rm gq}^{\,(3)}(x)$ to the quark-to-gluon splitting 
function in perturbative QCD. These moments, obtained by analytically
computing off-shell operator matrix elements for a general gauge group, 
agree with all known results, in particular with the moments $N\le 10$ 
derived before from structure functions in deep-inelastic scattering.  
Using the new moments and the available endpoint constraints, we 
construct approximations for $P_{\rm gq}^{\,(3)}(x)$ which improve upon 
those obtained from the lowest five even moments. The remaining 
uncertainties of this function are now practically irrelevant at 
momentum fractions $x > 0.1$. The resulting errors of the convolution of
$P_{\rm gq}$ at N$^3$LO with a typical quark distribution are small at 
$x \gsim 10^{-3}$ and exceed 1\% only at $x \lsim 10^{-4}$ for a strong 
coupling $\als = 0.2$. The present results for $P_{\rm gq}^{\,(3)}(x)$ 
should thus be sufficient for most collider-physics applications. 
\vspace*{0.5cm}
\end{titlepage}


Next-to-next-to-next-to-leading order (N$^3$LO) radiative corrections 
in QCD form an essential ingredient for precision physics at the 
Large Hadron Collider (LHC) \cite{Caola:2022ayt} and the future 
Electron-Ion Collider \cite{AbdulKhalek:2021gbh}. 
Almost a decade after the pioneering computation of the total cross 
section for Higgs-boson production via gluon-gluon fusion
\cite{Anastasiou:2015vya}, N$^3$LO partonic cross sections are 
becoming available for an increasing number of LHC processes.
Besides the partonic cross sections, fully consistent N$^3$LO 
calculations of collider-physics observables require the fourth-order 
contributions to the splitting functions for the scale dependence 
(evolution) of the parton distribution functions (PDFs). 
These N$^3$LO splitting functions are not fully known yet, but 
substantial progress towards their determination has been made over 
the past years. 

The N$^3$LO flavour non-singlet quark splitting functions 
$P_{\rm ns}^{\,(3)}(x)$ are fully known in the limit of a large 
number of colours $n_c$ \cite{Moch:2017uml}. 
Analytic results for their $x$-dependence in full QCD are known for 
the leading \cite{Gracey:1994nn} and next-to-leading 
\cite{Davies:2016jie} contributions in the limit of a large number of 
flavours~$\nf$; recently the QED-like $\cft \nf$ contribution has 
also been completed \cite{Gehrmann:2023iah}.
For all other colour factors, only the first eight even- or odd-$N$ 
moments have been published \cite{Moch:2017uml}; in the meantime 
those calculations have reached $N=22$ \cite{MVV-tba}.

So far, the situation is less favourable for the evolution of the 
flavour-singlet PDFs,
\beq
\label{eq:sgEvol}
 \frac{d}{d \ln\mus} \;
 \Big( \begin{array}{c}
         \! q_{\rm s}^{} \!\! \\ \!g\! \end{array} \Big)
 \: = \: \left( \begin{array}{cc}
         \! P_{\rm qq} & P_{\rm qg} \!\!\! \\
         \! P_{\rm gq} & P_{\rm gg} \!\!\! \end{array} \right)
 \otimes
 \Big( \begin{array}{c}
         \! q_{\rm s}^{}\!\! \\ \!g\!  \end{array} \Big)
 \quad \mbox{with} \quad
 P_{\,\rm ik}^{}(x,\als)
    \,=\, \sum_{n=0} \ar(n+1) P_{\,\rm ik}^{\,(n)}(x)
 \: ,
\eeq
where $a_{\rm s} = \als/(4\pi) = g_s^2/(4\pi)^2$ denote the strong 
coupling, $q_{\rm s}^{} \,=\, \sum_{\,i=1}^{\,\nf} \, 
( q_i^{} + \bar{q}_i^{} )$ and $g$ are the singlet quark and gluon 
PDFs, and $\otimes$ represents the Mellin convolution in the 
momentum variable~$x$. 
Only~the $\nft$ leading large-$\nf$ part of the splitting-function 
matrix in eq.~(\ref{eq:sgEvol}) is fully known at N$^3$LO 
\cite{Davies:2016jie}; 
recently the $\nfs$ contributions to the initial-quark quantities
$P_{\rm ps}^{\,(3)}(x)= P_{\rm qq}^{\,(3)}(x) - P_{\rm ns}^{\,(3)+}(x)$
and $P_{\rm gq}^{\,(3)}(x)$ have been derived 
in refs.~\cite{Gehrmann:2023cqm,Falcioni:2023tzp}. 
The even moments of $P_{\rm ik}^{\,(3)}(x)$ have been computed to 
$N=12$ for ${\rm ik} = {\rm qq}$ and to $N=10$ for all other entries 
in eq.~(\ref{eq:sgEvol}) using structure functions in deep-inelastic 
scattering (DIS) \cite{Moch:2021qrk,Moch:2023tdj}.

Up to a conventional sign, these even-$N$ moments are identical to 
the anomalous dimensions of the twist-2 operators of spin $N$, 
\beq
\label{eq:gamExp}
  \gamma_{\,\rm ik}^{}(N) 
  \:=\: -\int_0^1 \!dx\:x^{\,N-1}\,P_{\,\rm ik}^{}(x) 
\;, \quad 
  \gamma_{\,\rm ik}^{}(N,\als) 
  \:=\: \sum_{n=0} \ar(n+1)\,\gamma_{\,\rm ik}^{\,(n)}(N)
\; ,
\eeq
which can be computed efficiently by renormalizing off-shell operator 
matrix elements (OMEs) ${\rm A}_{\rm ik} 
 = \langle \,{\rm k}(p) | O_{\rm i} |\,{\rm k}(p)\rangle$,
where $O_{\rm i}$ indicates the quark ($\rm i= q$) or gluon 
($\rm i = g$) twist-2 operators. 
$\gamma_{\,\rm ps}^{\,(3)}$~and $\gamma_{\,\rm qg}^{\,(3)}$ 
were recently determined in this manner up to $N=20$ in 
refs.~\cite{Falcioni:2023luc,Falcioni:2023vqq}. Here we report on 
the corresponding results for $\gamma_{\,\rm gq}^{\,(3)}$.

In the non-abelian flavour-singlet cases,
the renormalisation of the OMEs involves a mixing between gauge 
invariant and unphysical operators, also known as {\it aliens} 
\cite{Dixon:1974ss,Joglekar:1975nu,Kluberg-Stern:1975ebk,%
Hamberg:1991qt,Falcioni:2022fdm,Gehrmann:2023ksf}. 
Following~ref.~\cite{Falcioni:2022fdm}, the aliens entering the 
calculation of $\gamma_{\,\rm gq}$ were organised as follows. 
The first class is given by       
\beq
\label{eq:alienI}
 O_{\!\rm A}^{\,I} \;=\; 
    \eta(N)\, \left(D.F^a + \gs \overline{\psi}\,\slash{\Delta}\,t^a 
    \psi\right)\, \left(\partial^{\,N-2}A_a \right)
\, , \quad
 O_{\rm c}^{\,I} \;=\;
    - \eta(N)\, \left(\partial \,\bar{c}^{\,a}\right) 
      \left(\partial^{\,N-1}c_a \right)
\, ,
\eeq
where $\Delta$ is a lightlike vector, $A_a= \Delta_\mu\, A_a^\mu\,$, 
$\partial = \Delta_\mu\,\partial^{\,\mu}$ and $D.F^a = \Delta_\nu 
(\partial_{\mu}\delta^{\,ab}+ \gs f^{acb}A^c_\mu)\,F^{\mu\nu,b}$. 
We~have determined $\eta(N\,\leq 20)$ to order $\as(3)$
and found agreement with ref.~\cite{Gehrmann:2023ksf}. 
Next we need
\bea
\label{eq:alienII}
 O_{\!\rm A}^{\,II} &\!=\!& 
 \gs\,f^{\,a\,a_1a_2} \left(D.F^a + \gs \overline{\psi}\,\slash{\Delta}\,
   t^a \psi\right) \sum_{\substack{n_1,\,n_2\,\geq\, 0\\n_1+n_2\,=\,N-3}}
   \kappa_{\,n_1n_2}^{(1)} \left(\partial^{\,n_1}A_{a_1} \right)
   \left(\partial^{\,n_2}A_{a_2} \right)
\, , \\[1mm] 
 O_{\rm c}^{\,II} &\!=\!&
  - \gs\,f^{\,a\,a_1a_2}\left(\partial \,\bar{c}^{\,a}\right) 
  \sum_{\substack{n_1,\,n_2\,\geq\, 0\\n_1+n_2\,=\,N-3}}
      \eta^{(1)}_{\,n_1 n_2}\left(\partial^{\,n_1}A_{a_1}\right)
      \left(\partial^{\,n_2+1}c_{a_2} \right)
\, .
\eea
 
\vspace*{-4mm}
\noindent
After picking a basis of independent constants $\eta^{(1)}_{\,n_1 n_2}$ 
and $\kappa^{\,(1)}_{\,n_1n_2}$ under the relations imposed by BRST and 
antiBRST symmetry \cite{Falcioni:2022fdm}, we have determined all the 
required two-loop mixing constants up to $N=20$.
The calculation of the renormalisation constants will be described
in detail in a future publication. 
We found complete agreement for those renormalisation constants which
were given in ref.~\cite{Gehrmann:2023ksf}. 
Furthermore we require the alien operators
\bea
\label{eq:alienIII}
 O_{\!\rm A_1}^{\,III} &\!=\!\! & 
  \gss f^{\,a\,a_1x}f^{\,a\,_2a_3x}\,
  \left(D.F^a + \gs \bar{\psi}\slash{\Delta}T^a\psi\right) \!\!\!
  \sum_{\substack{n_1,\,n_2,\,n_3\,\geq\,0\\n_1+n_2+n_3\,=\,N-4}} \!\!\!
  \kappa^{(1)}_{\,n_1 n_2 n_3}\,\left(\partial^{\, n_1}A_{a_1}\right)
  \left(\partial^{\,n_2}A_{a_2}\right)
  \left(\partial^{\,n_3}A_{a_3}\right)
\,, \;\; \\ 
O_{\!\rm A_2}^{\,III} &\!=\!\! & 
  \gss d^{\,a\,a_1a_2a_3}\,
  \left(D.F^a + \gs \bar{\psi}\slash{\Delta}T^a\psi \right) \!\!
  \sum_{\substack{n_1,\,n_2,\,n_3\,\geq\,0\\n_1+n_2+n_3\,=\,N-4}} \!\!
  \kappa^{(2)}_{\,n_1 n_2 n_3}\,\left(\partial^{\, n_1}A_{a_1}\right)
  \left(\partial^{\,n_2}A_{a_2}\right)\left(\partial^{\,n_3}A_{a_3}\right)
\,,\\ 
 O_{\rm c}^{\,III} &\!=\!\!	&
  -\gss f^{\,a\,a_1x}f^{\,a\,_2a_3x}\,\left(\partial\bar{c}^{\,a}\right)
  \!\! \sum_{\substack{n_1,\,n_2,\,n_3\,\geq\,0\\n_1+n_2+n_3\,=\,N-4}} 
  \!\! \eta^{(1)}_{\,n_1 n_2 n_3}\,\left(\partial^{\,n_1}A_{a_1}\right)
  \left(\partial^{\,n_2}A_{a_2}\right)\left(\partial^{\,n_3+1}c_{a_3}
  \right)
\, .
\eea
 
\vspace*{-4mm}
\noindent
We have computed a basis of independent mixing constants 
$\kappa_{\,ijk}^{\,(1)}$ at one-loop order, finding agreement with 
ref.~\cite{Gehrmann:2023ksf}. 
While the gluonic and ghost parts were already presented in 
ref.~\cite{Falcioni:2022fdm}, the quark contributions to the operators
in eq.~(\ref{eq:alienIII}) are presented here for the first time in 
this form. 

The calculation of $\gamma_{\,\rm gq}^{\,(3)}$ requires the OMEs 
$A_{\rm gq}$ to four loops and $A_{\rm qq}$ to three loops. 
In addition, the OMEs $A_{\rm iq}$ for all alien operators i are needed 
at the third order. 
The Feynman diagrams for the OMEs have been generated using {\sc Qgraf}
\cite{Nogueira:1991ex} and processed, see ref.~\cite{Herzog:2016qas} 
using a {\sc Form} \cite{Vermaseren:2000nd,Kuipers:2012rf,Ruijl:2017dtg}
program that classifies them according to their colour factors 
\cite{vanRitbergen:1998pn} and topologies. 
The calculation of the two-point functions has been performed by an 
optimized
in-house version of {\sc Forcer} \cite{Ruijl:2017cxj}. 
By renormalizing the OMEs computed in $4-2\ep$ dimensions, we have 
obtained the anomalous dimension $\gamma_{\,\rm gq}^{\,(3)}$. For QCD, 
i.e., the gauge group SU($n_c\!=\!3$), this results in the numerical 
values
\bea
\gamma_{\,\rm gq}^{(3)}(N\!=\!2) \; & =\!& 
  - 16663.2255 
  + 4439.14375 \,\nf 
  - 202.555479 \,\nfs
  - 6.37539072 \,\nft
\: , \nn \\[-0.5mm]
\gamma_{\,\rm gq}^{(3)}(N\!=\!4) \; & =\!& 
  - 6565.75315 
  + 1291.00675 \,\nf 
  - 16.1461902 \,\nfs 
  - 0.83976340 \,\nft
\: , \nn  \\[-0.5mm]
\gamma_{\,\rm gq}^{(3)}(N\!=\!6) \; & =\!& 
  - 3937.47937 
  + 679.718506 \,\nf 
  - 1.37207753 \,\nfs 
  - 0.13979433 \,\nft
\: , \nn  \\[-0.5mm]
\gamma_{\,\rm gq}^{(3)}(N\!=\!8) \; & =\!& 
  - 2803.64411 
  + 436.393057 \,\nf 
  + 1.81494625 \,\nfs 
  + 0.07358858 \,\nft
\: , \nn  \\[-0.5mm]
\gamma_{\,\rm gq}^{(3)}(N\!=\!10) &\!=\!& 
  - 2179.48761 
  + 310.063163 \,\nf
  + 2.65636842 \,\nfs
  + 0.15719522 \,\nft
\: , \nn  \\[-0.5mm]
\gamma_{\,\rm gq}^{(3)}(N\!=\!12) &\!=\!& 
  - 1786.31231 
  + 234.383019 \,\nf
  + 2.82817592 \,\nfs
  + 0.19211953 \,\nft
\: , \nn  \\[-0.5mm]
\gamma_{\,\rm gq}^{(3)}(N\!=\!14) &\!=\!& 
  - 1516.59810 
  + 184.745296 \,\nf
  + 2.78076831 \,\nfs
  + 0.20536518 \,\nft
\: , \nn  \\[-0.5mm]
\gamma_{\,\rm gq}^{(3)}(N\!=\!16) &\!=\!& 
  - 1320.36106 
  + 150.076970 \,\nf
  + 2.66194730 \,\nfs
  + 0.20798493 \,\nft
\: , \nn  \\[-0.5mm]
\gamma_{\,\rm gq}^{(3)}(N\!=\!18) &\!=\!& 
  - 1171.29329 
  + 124.717778 \,\nf
  + 2.52563073 \,\nfs
  + 0.20512226 \,\nft
\: , \nn  \\[-0.5mm]
\gamma_{\,\rm gq}^{(3)}(N\!=\!20) &\!=\!& 
  - 1054.26140 
  + 105.497994 \,\nf
  + 2.39223358 \,\nfs
  + 0.19938504 \,\nft
\: .
\label{eq:N20SU3}
\eea
The $\nfs$ and $\nft$ parts of eqs.~(\ref{eq:N20SU3}) agree with the
all-$N$ expressions of refs.~\cite{Davies:2016jie,Falcioni:2023tzp}.
The values at $N \leq 10$ agree with those obtained in refs.~\cite
{Moch:2021qrk,Moch:2023tdj} from structure functions in DIS -- a 
conceptionally much simpler but computationally much more involved 
approach.  
Our new exact results at $N \geq 12$, expressed in terms of rational 
numbers and values of Riemann's $\zeta$-function, can be found in 
eqs.~(\ref{Ggq3N12}) -- (\ref{Ggq3N20}) in the appendix for a general 
compact simple gauge group.

The all-$N$ expressions of the N$^n$LO anomalous dimensions in the
\MSb\ scheme include $\zeta_m$ up to $m_{\rm max}^{} = 2n-1$ 
(but not $\z2$), powers of simple denominators $D_a$ and harmonic 
sums $S_{\vec{w}}(N)$ \cite{Vermaseren:1998uu}, see also ref.~\cite
{Blumlein:1998if}.  
Up to three loops (N$^2$LO) \cite{Moch:2004pa,Vogt:2004mw} the 
anomalous dimensions (\ref{eq:gamExp}) have the form
\bea
\label{eq:Gij012N}
  \gamma_{\,\rm ik}^{\,(n)}(N)\big|_{\zeta_{\:\!m}} \!&\!=\!&
  \sum_{\;a^{\phantom{a}}} \; \sum_{p=p_0}^{2n+1-m} \:\: 
  \sum_{w=0}^{2n+1-m-p} \!\! c_{mapw}^{\,(n)}\, D_a^{\,p}\, S_w(N)
\quad \mbox{where } \quad D_a \;=\; \frac{1}{(N+a)} \; ,
\\[-1mm]
\label{eq:Hsums}
  S_{\pm m_1^{}}(N) 
  \; &=\!& \sum_{j=1}^{N}\; (\pm 1)^j \, \frac{1}{j^{\,m_1^{}}}
\; , \quad
  S_{\pm m_1^{},\,m_2^{}}(N) \;=\; \sum_{j=1}^{N}\:
  (\pm 1)^{j} \; \frac{1}{j^{\,m_1^{}}}\; S_{m_2^{}}(j)
\; , \quad \ldots
\eea
with $\z0 \equiv 1$, offsets $a = -1,\,0,\,1,\,2$ and minimal 
powers $p_0^{} = 0$ for i$\,=\,$k and $p_0^{} = 1$ for i$\,\neq\,$k. 
$S_w(N)$ in eq.~(\ref{eq:Gij012N}) is a shorthand for all 
$2^w\!-\!1$ harmonic sums of a given weight $w$ -- defined as the 
sum of the absolute values of the indices~$m_i$ ($m_i = -1$ does not 
occur) -- with $S_0(N) \equiv 1$. The coefficients $c_{mapw}^{\,(n)}$ 
are integer, modulo small-prime denominators that can be readily 
removed.  Thus Diophantine equations can be used for their 
determination from a limited number of $N$-values.

On top of the above contributions, terms with $D_{-2}$ (in special
combinations with sums $S_{\vec{w}}$, as there is no pole at $N=2$)
occur in the coefficient functions for inclusive DIS already at
two loops \cite{vanNeerven:1991nn,Zijlstra:1991qc,Moch:1999eb}.  
At the third order in $\als$, positive powers of $N$ enter the DIS 
coefficient functions with a special weight-5 combination of sums 
and Riemann-$\zeta$ values \cite{Vermaseren:2005qc},
\beq
\label{fNfct}
     f(N) \;=\;
        5 \* \zr5
       - 2 \* \S(-5)
       + 4 \:\!\* \zr3 \:\!\* \S(-2)
       - 4 \:\!\* \Ss(-2,-3)
       + 8 \* \Sss(-2,-2,1)
       + 4 \:\!\* \Ss(3,-2)
       - 4 \:\!\* \Ss(4,1)
       + 2 \* \S(5)
\; .
\eeq
If such additional contributions are present in $\gamma_{\,\rm ik}(N)$
beyond three loops, then they are visible already in the terms with 
$\zeta_{\:\!m \geq 3}$, in the case of $D_{-2}$ by the appearance of 
delta-function contributions $\ddelta_(N-2)$.
The comparatively simple $\zeta_{\:\!m \geq 3}$ terms can thus provide 
useful information about the most complex parts of the functions, 
the non-$\zeta$ contributions.
 
Up to terms in addition to eq.~(\ref{eq:Gij012N}), we expect the 
$\z5$ contributions to $\gamma_{\,\rm gq}^{\,(3)}$ to resemble the $\z3$
parts of $\gamma_{\,\rm gq}^{\,(2)}$. 
In particular, $S_1 \equiv S_1(N)$ should enter with the leading-order 
combination of $1/(N\!+\!a)$ denominators, 
$\pgq \,=\, 2/(N\!-\!1) - 2/N + 1/(N\!+\!1)$.
The available 10 moments are then sufficient for a direct determination 
of the coefficients $c_{5apw}^{\,(3)}$ by systems of linear equations, 
resulting in 
\def\col{\color{blue}}
\bea
\label{eq:Ggq3z5N}
 && \hspn \gamma_{\,\rm gq}^{\,(3)}(N)\big|_{\z5} =
%
   160\,\* \cff \* \Big(
       (N+2) / 12
     - 28\,\* \D(-1)
     - 239/3\,\* \D(0)
     + 473/6\,\* \D(1)
     + 52\,\* \Dd(0,2)
     + 26\,\* \Dd(1,2)
     + 14\,\* \pgq \,\* \S(1)
   \Big)
\nn \\[-0.5mm] & & \mbox{}
   + 160\,\* \cft \*\ca\,\* \Big(
     - (N+2) / 6
     + 44\,\* \D(-1)
     + 442/3\,\* \D(0)
     - 428/3\,\* \D(1)
     - 92\,\* \Dd(0,2)
     - 46\,\* \Dd(1,2)
     - 22\,\* \pgq \,\* \S(1)
   \Big)
\nn \\[-0.5mm] & & \mbox{}
  + 80\,\* \cfs\*\cas\,\* \Big(
       (N+2) / 6
     - 44\,\* \D(-1)
     - 548/3\,\* \D(0)
     + 1007/6\,\* \D(1)
     - 8/3\,\* \D(2)
     + 102\,\* \Dd(0,2)
     + 51\,\* \Dd(1,2)
\nn \\[-0.5mm] & & \mbox{}
     - 8\,\* \Dd(-1,2)
     + 25\,\* \pgq \,\* \S(1)
   \Big)
  + 80/3\,\* \cf\*\cat\,\* \Big(
       26\,\* \D(-1)
     + 224/3\,\* \D(0)
     - 166/3\,\* \D(1)
     + 16/3\,\* \D(2)
     - 24\,\* \Dd(0,2)
\nn \\[-0.5mm] & & \mbox{}
     - 12\,\* \Dd(1,2)
     + 16\,\* \Dd(-1,2)
     - 19\,\* \pgq \,\* \S(1)
   \Big)
  + 320\,\*\dfRAnR\*\ \Big(
     - 10 \,\* \D(-1)
     - 202/3 \,\* \D(0)
     + 391/6 \,\* \D(1)
\nn \\[-0.5mm] & & \mbox{}
     - 8/3 \,\* \D(2)
     - 8 \,\* \Dd(-1,2)
     + 42 \,\* \Dd(0,2)
     + 21 \,\* \Dd(1,2)
     + 8 \,\* \pgq \,\* \S(1)
   \Big)
  + 320\,\*\nf\,\*\cft\,\* \pgq
  - 160/3\,\*\nf\,\*\cfs\*\ca\,\* \pgq
\nn \\[-0.5mm] & & \mbox{}
  + 160/9\,\*\nf\,\*\cf\*\cas\,\* \Big(
     - 26\,\* \D(-1)
     + 62\,\* \D(0)
     - 49\,\* \D(1)
     - 24\,\* \Dd(0,2)
     - 12\,\* \Dd(1,2)
   \Big)
\nn \\[-0.5mm] & & \mbox{}
  + 640/3\,\*\nf\,\*\dfRRnR\*\ \Big(
       64 \,\* \D(0)
     - 68 \,\* \D(1)
     +  8 \,\* \D(-1)
     - 48 \,\* \Dd(0,2)
     - 24 \,\* \Dd(1,2)
   \Big)
\; .
\eea
\def\col{\color{black}}
\noindent
The contributions of the two quartic colour factors, 
see eqs.~(\ref{eq:d4def}) and (\ref{eq:d4fxSUn}), have been obtained 
before \cite{Moch:2018wjh}, the remaining terms are new. 
The $C_F^{\,k}\,C_A^{\,4-k}$ contributions at $k\geq 2$ indeed include 
terms beyond eq.~(\ref{eq:Gij012N}); these numerator-$(N\!+\!2)$ terms
hint at the appearance of the function $(N\!+\!2) f(N)$ in the 
$\nfz$-part of $\gamma_{\,\rm gq}^{\,(3)}$. This situation is completely
analogous to $\gamma_{\,\rm qg}^{\,(3)}$ in ref.~\cite{Falcioni:2023vqq},
where these terms are of the form $(N\!-\!1) \z5$, cf.~the relation 
between the off-diagonal $\gamma_{\,\rm ik}^{\,(3)}$ in 
ref.~\cite{Moch:2018wjh}.
In both cases these extra contributions vanish for $C_F = C_A$, which 
is part of the choice of the colour factors that leads to a 
${\cal N}\! =\! 1$ supersymmetric theory, for lower-order discussions
see refs.~\cite{Antoniadis:1981zv,Almasy:2011eq}.

The $\z4$ parts of $\gamma_{\,\rm ik}$ are special, as all 
$\pi^{\:\!2}$ terms -- see ref.~\cite{Herzog:2018kwj} for first 
five-loop results that include also $\z6 \propto \pi^{\:\!6}$.
Using the three-loop DIS coefficient functions of 
refs.~\cite{Vermaseren:2005qc,Soar:2009yh},
the $\z4$ part of $\gamma_{\,\rm gq}^{\,(3)}$ can be predicted
from the no-$\pi^{\,2}$ theorem 
\cite{Jamin:2017mul,Baikov:2018wgs,Kotikov:2019bqo}; 
the result is given by eq.~(11) of ref.~\cite{Davies:2017hyl}.

Ten $N$-values are not sufficient for all-$N$ determinations of the 
more complex $\nfz\,\z3$ and $\nfo\,\z3$ contributions to 
$\gamma_{\,\rm gq}^{\,(3)}$ -- except for the quartic colour factors,
which are very closely related \cite{Moch:2018wjh} to those occurring in
$\gamma_{\,\rm qg}^{\,(3)}$ \cite{Falcioni:2023vqq}. 
For one colour factor, $\nf\,\cft$, the diagrams are simple enough to 
allow the extension of our {\sc Forcer} calculations to $N=30$. 
This was sufficient to obtain and check the corresponding
coefficients $c_{3apw}^{\,(3)}$ in eq.~(\ref{eq:Gij012N}) by a system
of Diophantine equations. We hence find
\def\col{\color{blue}}
\bea
\label{eq:Ggq3z3N}
&& \hspn \gamma_{\,\rm gq}(N)\big|_{\z3} \;=\; 
%
   128\,\* \dfRAnR\,\*\Big( \big\{
         7/12
       + 33/2\,\* \eta
       - 96\,\* \S(1)\,\* \eta 
       + 32\,\* \S(1)\,\* \nu
       - 24\,\* \S(-2)
       - 6\,\* \S(1)\,\* \eta^2
\nn \\ & & \mbox{}
       + 10\,\* \S(-2)\,\* \eta 
       - 16\,\* \S(-2)\,\* \nu
       + 43\,\* \Sp(1,2)\,\* \eta 
       - 16\,\* \Sp(1,2)\,\* \nu
       - 8\,\* (\S(-2,1)-\S(1,-2))
       + 8\,\* \S(1)\,\*\S(-2)
       - \S(3)
	 \big\}\,\*\pgq
\nn \\[0.5mm] & & \mbox{}
     - \big\{
         40
       + 24\,\* \S(1)
       - 72\,\* \S(-2)
       - 10\,\* \Sp(1,2)
       + 72\,\* (\S(-2,1)-\S(1,-2))
       + 72\,\* \S(1)\,\*\S(-2)
       + 36\,\* \S(3)
     \big\}			
     \,\*\D(-1)\,\*\D(0)\,\*\D(1)
   \Big)
\nn \\ & & \mbox{}
   + 32/9\,\*\nf\,\*\cft\,\* \Big(
       - 10/3\,\* \ddelta_(N-2)
       - 1372/3\,\*\D(-1)
       - 2233/2\,\*\D(0)
       + 6049/4\,\*\D(1)
       - 20/3\,\*\D(2)
\nn \\ & & \mbox{}
       - 16\,\*\Dd(-1,2)
       + 1477\,\*\Dd(0,2)
       + 305\,\*\Dd(1,2)
       + 32\,\*\Dd(2,2)
       - 186 \,\*\Dd(0,3)
       - 87\,\*\Dd(1,3)
       + 216\,\* (\Dd(0,4) - \Dd(1,4))
     - \big\{
         30\,\*\D(-1)
\nn \\[1.5mm] & & \mbox{}
       - 856\,\*\D(0)
       + 725\,\*\D(1)
       - 32\,\*\D(2)
       - 96\,\*\Dd(-1,2)
       + 288\,\*\Dd(0,2)
       + 300\,\*\Dd(1,2)
       - 108\,\* (2\,\*\Dd(0,3) - \Dd(1,3))
       \big\} \,\* \S(1)
\nn \\ & & \mbox{}
       + \big\{
         60\,\* \S(-2) 
       - 78\,\* \S(1,1)      
       + 72\,\* \S(2)     
       \big\} \,\*\pgq \Big)
 + 256\,\* \nf\,\*\dfRRnR\,\*\Big(
       \big\{
       - 7/12
       - 8\,\* \S(1)\,\* \eta^2
       + 8\,\* \S(-2)\,\* \eta
\nn \\ & & \mbox{}
       - 3\,\* \eta
       \big\} \,\*\pgq
     + \big\{
          4
       - 16\,\* \S(-2)
       \big\}
       \,\*\D(-1)\,\*\D(0)\,\*\D(1)
       \Big)
  + 64/27\,\* \nfs\,\* \cfs\,\* \Big(
         52\,\*\D(-1)
       - 89\,\*\D(0)
       + 76\,\*\D(1)
       + 4\,\*\D(2)
\; \nn \\ & & \mbox{}
       + 12\,\*\Dd(-1,2)
       + 42\,\*\Dd(0,2)
       + 9\,\*\Dd(1,2)
       - 6\,\* \S(1) \,\* \pgq
    \Big)
  + 32/9\,\* \nfs\,\* \cf\*\ca\,\* \Big(
         \ddelta_(N-2) / 3
       - 103/3\,\*\D(-1)
       + 63\,\*\D(0)
\nn \\ & & \mbox{}
       - 54\,\*\D(1)
       - 8/3\,\*\D(2)
       - 8\,\*\Dd(-1,2)
       - 28\,\*\Dd(0,2)
       - 12\,\*\Dd(1,2)
       + 6\,\* \S(1) \,\* \pgq
     \Big)
 - 128/27\,\* \nft\*\,\cf\,\* \pgq 
 + \, \ldots 
\eea
\def\col{\color{black}}
with
\beq
  \pgq \;=\; 2 \D(-1) - 2 \D(0) + \D(1)
\;, \quad
  \eta  \;=\; \D(0) - \D(1)
\;, \quad
  \nu  \;=\; \D(-1) - \D(2)
\; .
\eeq
As above, we have omitted also the argument $N$ of the sums for 
brevity. A very considerable effort would be required to find the 
contributions, indicated by `$+\:\ldots$', that are still missing 
in eq.~(\ref{eq:Ggq3z3N}).

Also eq.~(\ref{eq:Ggq3z3N}) includes a structure that did not occur
in the anomalous dimensions up to the third order in $\als$ but 
was mentioned above: Both the $\nf \cft$ and $\nfs C_F C_A$ terms
receive a $\ddelta_(N\!-\!2)$ contribution. In the latter case, which
is fully known \cite{Falcioni:2023tzp}, this arises from the 
non-$\zeta$ contribution $(N\!-\!2)^{-1}\, S_{-2}(N\!-\!2)$, 
corresponding to 
$\,x^{\:\!-2}\, {\rm H}_{-1,0}(x) = 
 \,x^{\:\!-2}\, [ \,\ln(x) \ln(1+x) + \mbox{Li}_2(-x) ]$
\cite{Remiddi:1999ew,Moch:1999eb}.
Exactly the same function occurs in the two-loop coefficient functions
for inclusive DIS \cite{vanNeerven:1991nn,Zijlstra:1991qc,Moch:1999eb}
and in the $\nfs C_F C_A$ contribution to $\gamma_{\,\rm ps}^{\,(3)}(N)$
recently determined in ref.~\cite{Gehrmann:2023cqm}.

The above partial all-$N$ results are not relevant to N$^3$LO analyses 
of LHC processes. Until the complete functions $P_{\rm ik}^{\,(3)}(x)$ 
become known, such analyses will have to rely on approximations based
on the available moments and information about the large-$x$ 
(threshold) and small-$x$ (high-energy) limits.
With eqs.~(\ref{eq:N20SU3}) we are now in the position to improve upon 
the $N \leq 10$ based approximations of ref.~\cite{Moch:2023tdj}, thus 
putting $P_{\rm gq}^{\,(3)}$ on the same footing as 
$P_{\rm ps}^{\,(3)}$ and $P_{\rm qg}^{\,(3)}$ in refs.~\cite
{Falcioni:2023luc,Falcioni:2023vqq}.

Up to the presence of a leading-logarithmic BFKL small-$x$ contribution
\cite{Jaroszewicz:1982gr,CataniFM90}, the endpoint structure of 
$P_{\rm gq}^{\,(3)}(x)$ is the same as that of $P_{\rm qg}^{\,(3)}(x)$
given in eqs.~(10) and (11) of ref.~\cite{Falcioni:2023vqq}. 
So far the next-to-leading logarithmic (NLL) small-$x$ correction has 
been calculated \cite{Fadin:1998py,Ciafaloni:1998gs} and transformed 
to the \MSb\ scheme \cite{Ciafaloni:2005cg,Ciafaloni:2006yk} only for
$P_{\rm gg}$.
At order $\as(3)$, the NLL BFKL contributions to $P_{\rm gq}$ and 
$P_{\rm gg}$ are related by Casimir scaling in the large-$\nc$ limit. 
In QCD the breaking of this relation is numerically small, see 
eq.~(4.29) of ref.~\cite{Vogt:2004mw}.
In view of the large uncertainties due to the unknown NNLL
$x^{\,-1} \ln x$ terms, see below, no relevant bias should be generated
by assuming Casimir scaling for the $x^{\,-1} \ln^{\,2} x$ term of 
$P_{\rm gq}^{\,(3)}$.
Under this assumption the known endpoint contributions analogous to 
eq.~(15) of ref.~\cite{Falcioni:2023vqq} are, to eight significant 
figures, given by
\cite{Jaroszewicz:1982gr,CataniFM90,Fadin:1998py,Ciafaloni:1998gs,%
Ciafaloni:2005cg,Ciafaloni:2006yk,Davies:2022ofz,Soar:2009yh,%
Vogt:2010cv,Almasy:2010wn}
\bea
\label{eq:Pgq30-nf}
{\lefteqn{ \!\!
 p_{{\rm gq},0}^{\,(\nf)}(x) \;=\;
 - 3692.7188\,\*L_0^3/x
 - (47516.440 + 442.83691\,\*\nf)\,\* L_0^2/x
}}
\nn \\ && \mbox{}
 + (52.235940 - 7.3744856\,\*\nf)\,\* L_0^6
 - (292.21399 - 1.8436214\,\*\nf)\,\* L_0^5
\nn \\ && \mbox{}
 + (7310.6077 - 378.87135\,\*\nf - 32.438957\,\*\nfs)\,\* L_0^4
 + (13.443073 - 0.54869684\,\*\nf)\,\* L_1^5
\nn \\ && \mbox{}
 + (375.39831 - 34.494742\,\,\*\nf + 0.87791495\,\*\nfs)\,\*L_1^4
 + (22.222222 - 0.54869684\,\*\nf)\,\* x_1\* L_1^5
\nn \\ && \mbox{}
 + (662.42163 - 47.992684\,\*\nf + 0.87791495\,\*\nfs)\,\* x_1\* L_1^4
\; ,
\eea
where we have used the abbreviations $x_1^{}=1\!-\!x$, 
$L_1=\ln (1\!-\!x)$ and $L_0=\ln x$. The coefficient of $L_0^{\,2}/x$
in the first line of eq.~(\ref{eq:Pgq30-nf}) will be referred to as 
$\bfkl1$ below.

Our procedure for the construction of approximate expressions for
$P_{\rm gq}^{\,(3)}(x)$ is identical to that used for its counterpart
$P_{\rm qg}^{\,(3)}(x)$ in ref.~\cite{Falcioni:2023vqq}.
Also here it is not possible, even with ten moments, to make a `direct
fit' for the coefficient of $x^{\,-1} \ln x$. In fact, knowing five
additional moments on top of those employed in ref.~\cite{Moch:2023tdj}
leads to considerable improvements elsewhere, see below, but does not 
enable us to assign a smaller uncertainty to this important coefficient.

The approximations selected to represent our error bands for the 
physically relevant numbers $\nf = 3,\,4,\,5 $ of light flavours read
(with $\bfkl1$ as defined below eq.~(\ref{eq:Pgq30-nf}))
\bea
\label{eq:Pgq3A3-nf3}
{\lefteqn{
 P_{\rm gq,\,A}^{\,(3)}(\nf=3,x) \; = \;
 p_{{\rm gq},0}^{\,(\nf=3)}(x) +
             6\,\bfkl1\,\*L_0/x
           - 744384\,\*x_1/x
           + 2453640
           - 1540404\,x\*(2+x)
}}
\nn \\ && \mbox{\hspm}
           + 1933026\,\*L_0
           + 1142069\,\*L_0^2
           + 162196 \,\*L_0^3
           - 2172.1   \*L_1^3
           - 93264.1  \*L_1^2
           - 786973 \,\*L_1
           + 875383 \,\*x_1\*L_1^2
,
\nn\\[1mm]
{\lefteqn{
 P_{\rm gq,\,B}^{\,(3)}(\nf=3,x) \; = \;
 p_{{\rm gq},0}^{\,(\nf=3)}(x) +
             3\,\*\bfkl1\,\*L_0/x
           + 142414 \,\*x_1/x
           - 326525
           + 2159787\,x\*(2-x)
}}
\nn \\ && \mbox{\hspm}
           - 289064 \,\*L_0
           - 176358 \,\*L_0^2
           + 156541 \,\*L_0^3
           + 9016.5 \,\*L_1^3
           + 136063 \,\*L_1^2
           + 829482 \,\*L_1
           - 2359050\,\*L_0\*L_1
\, ,
\nn \\[-1mm]
\\[-3mm]
\label{eq:Pgq3A3-nf4}
{\lefteqn{
 P_{\rm gq,\,A}^{\,(3)}(\nf=4,x) \; = \;
 p_{{\rm gq},0}^{\,(\nf=4)}(x) +
             6\,\bfkl1\,\*L_0/x
           - 743535 \,\*x_1/x
           + 2125286
           - 1332472\,x\*(2+x)
}}
\nn \\ && \mbox{\hspm}
           + 1631173\,\*L_0
           + 1015255\,\*L_0^2
           + 42612  \,\*L_0^3
           - 1910.4 \,\*L_1^3
           - 80851  \,\*L_1^2
           - 680219 \,\*L_1
           + 752733 \,\*x_1\*L_1^2
\, ,
\nn\\[1mm]
{\lefteqn{
 P_{\rm gq,\,B}^{\,(3)}(\nf=4,x) \; = \;
 p_{{\rm gq},0}^{\,(\nf=4)}(x) +
             3\,\*\bfkl1\,\*L_0/x
           + 160568 \,\*x_1/x
           - 361207
           + 2048948\,x\*(2-x)
}}
\nn \\ && \mbox{\hspm}
           - 245963 \,\*L_0
           - 171312 \,\*L_0^2
           + 163099 \,\*L_0^3
           + 8132.2 \,\*L_1^3
           + 124425 \,\*L_1^2
           + 762435 \,\*L_1
           - 2193335\,\*L_0\*L_1
\, ,
\nn \\[-1mm]
\\[-3mm]
\label{eq:Pgq3A3-nf5}
{\lefteqn{
 P_{\rm gq,\,A}^{\,(3)}(\nf=5,x) \; = \;
 p_{{\rm gq},0}^{\,(\nf=5)}(x) +
             6\,\bfkl1\,\*L_0/x
           - 785864 \,\*x_1/x
           + 285034
           - 131648 \,x\*(2+x)
}}
\nn \\ && \mbox{\hspm}
           - 162840 \,\*L_0
           + 321220 \,\*L_0^2
           + 12688  \,\*L_0^3
           + 1423.4 \,\*L_1^3
           + 1278.9 \,\*L_1^2
           - 30919.9\,\*L_1
           + 47588  \,\*x_1\*L_1^2
\, ,
\nn\\[1mm]
{\lefteqn{
 P_{\rm gq,\,B}^{\,(3)}(\nf=5,x) \; = \;
 p_{{\rm gq},0}^{\,(\nf=5)}(x) +
             3\,\*\bfkl1\,\*L_0/x
           + 177094 \,\*x_1/x
           - 470694
           + 1348823\,x\*(2-x)
}}
\nn \\ && \mbox{\hspm}
           - 52985  \,\*L_0
           - 87354  \,\*L_0^2
           + 176885 \,\*L_0^3
           + 4748.8 \,\*L_1^3
           + 65811.9\,\*L_1^2
           + 396390 \,\*L_1
           - 1190212\,\*L_0\*L_1
\nn
\, .\\[-1mm]
\\[-7mm] \nn
\eea
At $N=22$ we predict, with the brackets indicating a conservative 
uncertainty of the last digit:
\bea
\label{eq:ggq3N22appr}
  - \gamma_{\,\rm gq}^{\,(3)}(N\!=\!22) \;=\;
   662.7981(3)\, , \:\:
   549.2876(3)\, , \:\:
   426.6266(2)
\quad \mbox{for} \quad \nf\,=\,3,4,5
\; .
\eea

\begin{figure}[p]
\vspace{-4mm}
\centerline{\epsfig{file=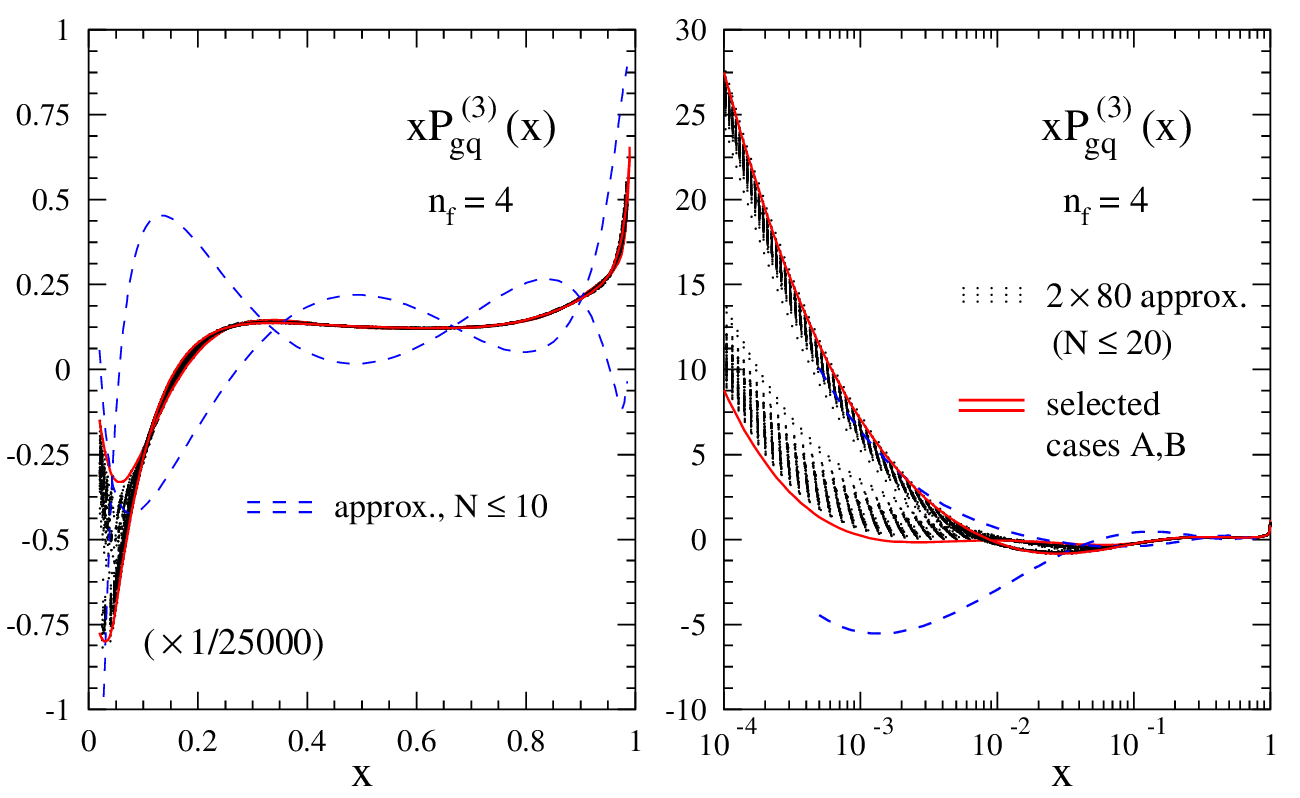,width=16.0cm,angle=0}}
\vspace{-3mm}
\caption{\label{fig:pgq3ab} \small
Two sets of 80 trial functions for the four-loop (N$^3$LO) 
contribution to the quark-to-gluon splitting function at $\nf = 4$.
The two cases selected for eq.~(\ref{eq:Pgq3A3-nf4}) are shown by
the solid (red) lines. Also shown, by the dashed (blue) lines, 
are the selected approximations of ref.~\cite{Moch:2023tdj} based 
on only the first 5 even moments.}
\end{figure}

The selection procedure and its results are illustrated in 
fig.~\ref{fig:pgq3ab} for $\nf = 4$; the two functions in 
eq.~(\ref{eq:Pgq3A3-nf5}) are shown by the red (solid) lines. 
A comparison with the corresponding $N\!\leq\! 10$ results of 
ref.~\cite{Moch:2023tdj}, shown by the blue (dashed) lines, 
reveals an impressive improvement down to $x \approx 10^{-2}$. 
$P_{\rm gq}^{\,(3)}(x)$ can now be considered as sufficiently 
well constrained at $x$-values above 0.1.

\begin{figure}[p]
\vspace{-2mm}
\centerline{\epsfig{file=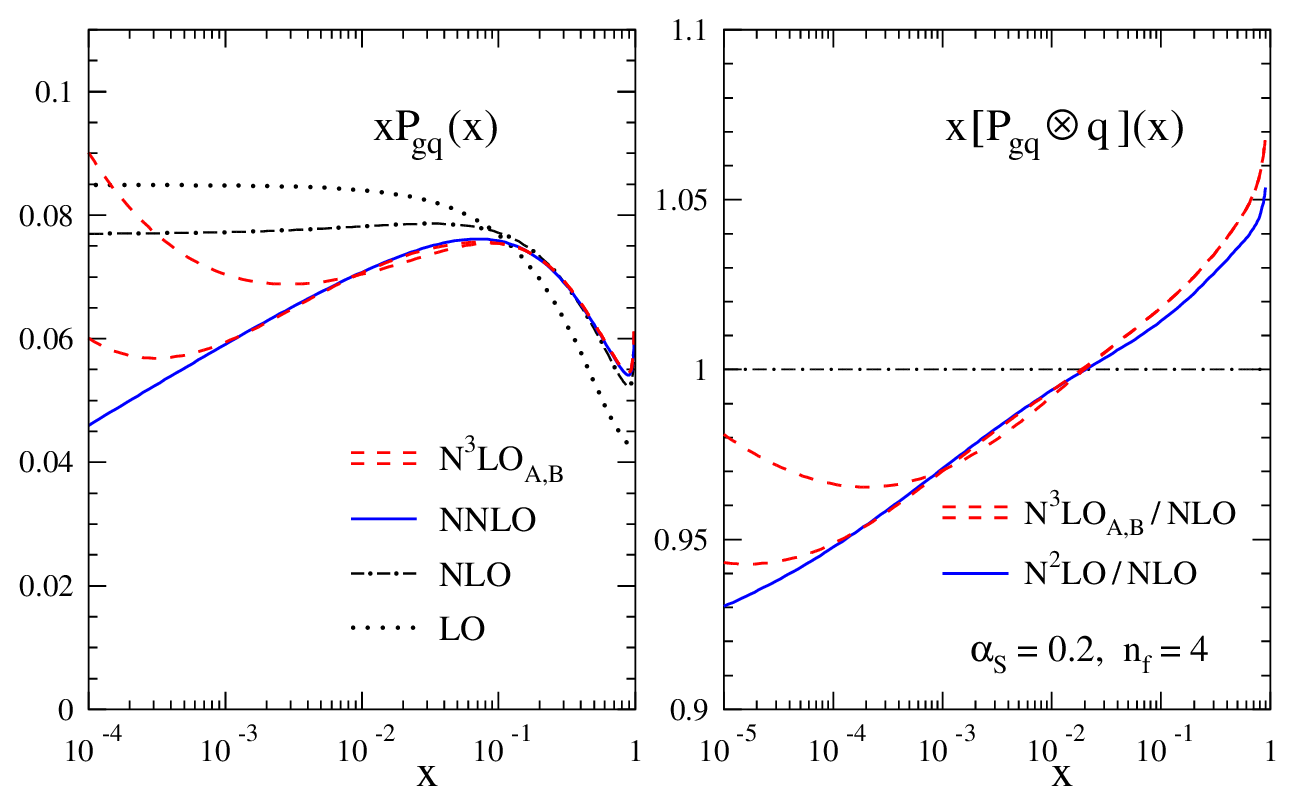,width=16.0cm,angle=0}}
\vspace{-2mm}
\caption{\label{fig:pgqn3lo} \small
Left: the perturbative expansion of the splitting functions
$P_{\rm gq}$ to N$^3$LO for $\nf = 4$ and $\als = 0.2$, using
eq.~(\ref{eq:Pgq3A3-nf4}) for the four-loop contribution.
Right: the resulting N$^2$LO and N$^3$LO convolutions 
with the reference quark distribution (\ref{eq:qs-shape})
normalized to the NLO results.}
\end{figure}

The uncertainties of $P_{\rm gq}^{\,(3)}(x)$ `as such' are still rather
large at $10^{-2} \lsim x < 10^{-1}$.
However, as shown in the left panel of fig.~\ref{fig:pgqn3lo},
the resulting uncertainties of the expansion of the splitting function
$P_{\rm gq}$ to N$^3$LO appear to be perfectly tolerable also in this
range, down to (for LHC purposes) rather low scales.
Of course, the splitting functions enter physical quantities only via
the convolution with the respective quark or gluon PDF.
This convolution is illustrated in the right panel of
fig.~\ref{fig:pgqn3lo} using a sufficiently realistic order-independent
model input \cite{Vogt:2004mw} for the singlet quark PDF in
eq.~(\ref{eq:sgEvol}),
\beq
\label{eq:qs-shape}
 xq_{\rm s}^{}(x,\mu_{0}^{\,2}) \; = \;
 0.6\, x^{\, -0.3} (1-x)^{3.5} \left(1 + 5.0 \,x^{\, 0.8}\right)
 \, ,
\eeq
together with a strong coupling of $\als (\mu_{0}^{\,2}) \,=\, 0.2$
corresponding to a scale $\mu_{0}^{\,2}\,\approx\, 30\ldots 50$ GeV$^2$.
Here the uncertainties of $P_{\rm gq}^{\,(3)}(x)$ do not appear to have
a relevant impact at $x \gsim 10^{-3}$.
An overall uncertainty of $\pm 1\%$ of the NLO result for $P_{\rm gq} 
\otimes q_{\rm s}^{}$ is reached only at $x \approx 10^{-4}$.
 
To summarize,
we have computed the even-$N$ moments $N \leq 20$ of the fourth-order
(N$^3$LO) quark-to-gluon splitting function $P_{\rm gq}^{\,(3)}(x)$ 
in the framework of the operator-product expansion.
These results facilitate the determination of the all-$N$ form of the
$\z5$ contribution and of parts of the corresponding $\z3$ 
terms. These expressions show deviations from the functional form of
the anomalous dimensions up to three loops which should be of interest 
to future research.

Together with (approximately) known endpoint constraints at $x \ra 1$
and $x \ra 0$, our additional moments -- only the results up to
$N = 10$ had been obtained before \cite{Moch:2021qrk,Moch:2023tdj} -- 
have been employed to construct improved approximations for 
$P_{\rm gq}^{\,(3)}(x)$ that should be sufficiently accurate for a 
wide range of phenomenological applications in collider physics.
Their uncertainties are still large at $x \lsim 10^{-3}$. A drastic
improvement in this region would be obtained if the next-to-leading
and next-to-next-leading small-$x$ contributions, $x^{-1} \ln^2 x$ 
and $x^{-1} \ln x$, became known.

%
\subsection*{Acknowledgements}
\vspace*{-1mm}
G.F.~\& S.M. are grateful to the Galileo Galilei Institute, Florence, 
for hospitality and support during the scientific program on {\it 
Theory Challenges in the Precision Era of the Large Hadron Collider},
where some work for this publication was done.
This work has been supported by the UKRI FLF Mr/S03479x/1 and the 
STFC Consolidated Grant ST/X000494/1;
the EU's Horizon Europe programme under the 
Marie Sklodowska-Curie grant 101104792, {\it QCDchallenge};
the DFG through the Research Unit FOR 2926,
project number 40824754, and DFG grant MO~1801/4-2,
the ERC Advanced Grant 101095857 {\it Conformal-EIC};
and the STFC Consolidated Grant ST/X000699/1.

{\small
\addtolength{\baselineskip}{-2.2mm}
\providecommand{\href}[2]{#2}\begingroup\raggedright\endgroup

}
\appendix
%
%
\renewcommand{\theequation}{\ref{sec:appA}.\arabic{equation}}
\setcounter{equation}{0}
\section{Mellin moments of $P^{\,(3)}_{\,\rm gq}$}
\label{sec:appA}

\vspace*{-3mm}
The exact results for the four-loop anomalous dimensions 
$\gamma^{\,(3)}_{\,\rm gq}(N)$ at even $2\leq N \leq 10$ have 
already been presented, for a general compact simple gauge group, 
in eqs.~(11) - (13) of ref.~\cite{Moch:2021qrk} and eqs.~(9) and (10) 
of ref.~\cite{Moch:2023tdj}. 
We agree with the results given in those articles, which were obtained 
in a different theoretical framework. Here we report the corresponding 
results at $12 \leq N \leq 20$. The~numerical values for QCD, i.e., 
SU($\nc=3$), have been given in eqs.~(\ref{eq:N20SU3}) above.

The quadratic Casimir invariants are $\ca= \nc$ and 
$\cf= (\ncs-1)/(2\nc)$ in SU$(\nc)$. 
The quartic group invariants are products of two symmetrized traces of 
four generators $T_r^a$ of the fundamental ($R$) or adjoint ($A$) 
representation,
\beq
\label{eq:d4def}
  d_{r}^{\,abcd} \; =\; \frac{1}{6}\: {\rm Tr} \, ( \, 
   T_{r}^{a\,} T_{r}^{b\,} T_{r}^{c\,} T_{r}^{d\,}
   + \,\mbox{ five $bcd$ permutations}\, ) 
\; ,
\eeq
 
\vspace*{-4mm}
\noindent
which leads to
\beq
\label{eq:d4fxSUn}
\dfRAnr \;=\;
  \frac{( \ncs -1 )\,( \ncs + 6 )}{48} 
\;\,, \quad
\dfRRnr \;=\;
  \frac{(\ncs - 1)\,( \ncf - 6\,\ncs + 18 )}{96\,\nct}
\; .
\eeq
Their values in QCD are 
$\dfRAnr \equiv \dfRAnR \,=\, 5/2$ and 
$\dfRRnr \equiv \dfRRnR \,=\, 5/36$.

\def\col{\color{blue}}

{\small{
\bea
&& \hspn \gamma_{\,\rm gq}^{\,(3)}(N=12) \,=\,
\cff\,\*\Big(-\frac{25682557611275387813242057003995677}{135916765937789612439776112000000}
-\frac{78873286642153277627}{313680161535741000}\,\*\z3
\nn \\[0.5mm] & & \mbox{\hspp}
+\frac{12335483}{193050}\,\*\z4
+\frac{17406716}{42471}\,\*\z5\Big)
+\ca\,\*\cft\,\*\Big(\frac{422699180143530974480654228020604813}{1143819017242696998194479488000000}
\nn \\[0.5mm] & & \mbox{\hspp}
+\frac{100064481514290063961}{313680161535741000}\,\*\z3
-\frac{760772056049}{9591882300}\,\*\z4
-\frac{237624788}{351351}\,\*\z5\Big)
\nn \\[0.5mm] & & \mbox{\hspp}
+\cas\,\*\cfs\,\*\Big(-\frac{224250321575347263949905289161744199}{1455769658308887088611155712000000}
-\frac{2010183037743051616}{39210020191967625}\,\*\z3
\nn \\[0.5mm] & & \mbox{\hspp}
+\frac{3810092749}{532882350}\,\*\z4
+\frac{1211419205}{3864861}\,\*\z5\Big)
+\cat\,\*\cf\,\*\Big(-\frac{605797324486893500585306535221627}{18906099458556975176768256000000}
\nn \\[0.5mm] & & \mbox{\hspp}
-\frac{1052537460264167687}{25347891841272000}\,\*\z3
+\frac{79289578229}{9591882300}\,\*\z4
-\frac{23156863}{1054053}\,\*\z5\Big)
\nn \\[0.5mm] & & \mbox{\hspp}
+\dfRAnr\,\*\Big(-\frac{113527295175322043699}{2082683928124800000}
-\frac{66259676094769}{321401840760}\,\*\z3
+\frac{1159790344}{3864861}\,\*\z5\Big)
\nn \\[0.5mm] & & \mbox{\hspp}
+\nf\,\*\cft\,\*\Big(-\frac{332043482084304364183967274124153}{19063650287378283303241324800000}
-\frac{393343213522261}{33948069430275}\,\*\z3
-\frac{6070628}{495495}\,\*\z4
\nn \\[0.5mm] & & \mbox{\hspp}
+\frac{12640}{429}\,\*\z5\Big)
+\nf\,\*\ca\,\*\cfs\,\*\Big(\frac{2936117949737136383600864384611}{157550828821308126473068800000}
-\frac{3457752318647119}{67896138860550}\,\*\z3
\nn \\[0.5mm] & & \mbox{\hspp}
+\frac{8619754063}{251215965}\,\*\z4
-\frac{6320}{1287}\,\*\z5\Big)
+\nf\,\*\cas\,\*\cf\,\*\Big(-\frac{3409052330654655340645389731}{1469005396935273906508800000}
\nn \\[0.5mm] & & \mbox{\hspp}
+\frac{25514383657902523}{407376833163300}\,\*\z3
-\frac{5541945667}{251215965}\,\*\z4
-\frac{119360}{5577}\,\*\z5\Big)
\nn \\[0.5mm] & & \mbox{\hspp}
+\nf\,\*\dfRRnr\,\*\Big(-\frac{14322664324006372519}{260335491015600000}
-\frac{33040603052}{2608781175}\,\*\z3
+\frac{421760}{5577}\,\*\z5\Big)
\nn \\[0.5mm] & & \mbox{\hspp}
+\nfs\,\*\cfs\,\*\Big(\frac{169033294355320918596314039}{79352523673735777985520000}
+\frac{81408784}{15810795}\,\*\z3
-\frac{5056}{1287}\,\*\z4\Big)
\nn \\[0.5mm] & & \mbox{\hspp}
+\nfs\,\*\ca\,\*\cf\,\*\Big(-\frac{1990546832712974636980331}{1109825505926374517280000}
-\frac{171649172}{57972915}\,\*\z3
+\frac{5056}{1287}\,\*\z4\Big)
\nn \\[0.5mm] & & \mbox{\hspp}
+\nft\,\*\cf\,\*\Big(\frac{2265689177355445577}{3387745744586002800}
-\frac{5056}{11583}\,\*\z3\Big)
\;, \label{Ggq3N12}
\\[2mm]
&& \hspn \gamma_{\,\rm gq}^{\,(3)}(N=14) \,=\,
\cff\,\*\Big(-\frac{1175858514183978179181088212329622671}{5838242900509599261617655720000000}
\nn \\[0.5mm] & & \mbox{\hspp}
-\frac{32053546295279752556}{124759155156260625}\,\*\z3
+\frac{10403997892}{165540375}\,\*\z4
+\frac{5234629744}{12297285}\,\*\z5\Big)
\nn \\[0.5mm] & & \mbox{\hspp}
+\ca\,\*\cft\,\*\Big(\frac{435029240281577148104394100536180709}{1077829458555618325221721056000000}
+\frac{172973892630128742931}{499036620625042500}\,\*\z3
\nn \\[0.5mm] & & \mbox{\hspp}
-\frac{5900838170593}{71016820875}\,\*\z4
-\frac{160919504}{223587}\,\*\z5\Big)
\nn \\[0.5mm] & & \mbox{\hspp}
+\cas\,\*\cfs\,\*\Big(-\frac{62684886023029557261923348044015457}{331639833401728715452837248000000}
-\frac{173609406082076933}{2268348275568375}\,\*\z3
\nn \\[0.5mm] & & \mbox{\hspp}
+\frac{231212408549}{15781515750}\,\*\z4
+\frac{4199718508}{12297285}\,\*\z5\Big)
+\cat\,\*\cf\,\*\Big(-\frac{4272807969344509309643357544788243}{255107564155175934963720960000000}
\nn \\[0.5mm] & & \mbox{\hspp}
-\frac{926544367976770183}{25591621570515000}\,\*\z3
+\frac{794134472909}{142033641750}\,\*\z4
-\frac{177252716}{7378371}\,\*\z5\Big)
\nn \\[0.5mm] & & \mbox{\hspp}
+\dfRAnr\,\*\Big(-\frac{21335213154567939744419}{414584269442343000000}
-\frac{613323018929398364}{3198952696314375}\,\*\z3
+\frac{3460252064}{12297285}\,\*\z5\Big)
\nn \\[0.5mm] & & \mbox{\hspp}
+\nf\,\*\cft\,\*\Big(-\frac{490379446041031040879323234108051}{40826873430137057773550040000000}
-\frac{33894392681638}{2403420508125}\,\*\z3
-\frac{46632848}{4099095}\,\*\z4
\nn \\[0.5mm] & & \mbox{\hspp}
+\frac{6784}{273}\,\*\z5\Big)
+\nf\,\*\ca\,\*\cfs\,\*\Big(\frac{40759120798908335841181047497491}{3140528725395158290273080000000}
-\frac{992731197367267}{26437625589375}\,\*\z3
\nn \\[0.5mm] & & \mbox{\hspp}
+\frac{65246478712}{2152024875}\,\*\z4
-\frac{3392}{819}\,\*\z5\Big)
+\nf\,\*\cas\,\*\cf\,\*\Big(-\frac{958430136662854610670607569547}{322105510296939311822880000000}
\nn \\[0.5mm] & & \mbox{\hspp}
+\frac{9023658476762282}{174488328889875}\,\*\z3
-\frac{40764233512}{2152024875}\,\*\z4
-\frac{4416}{245}\,\*\z5\Big)
\nn \\[0.5mm] & & \mbox{\hspp}
+\nf\,\*\dfRRnr\,\*\Big(-\frac{86703311715954607957}{1850822631439031250}
-\frac{244011856707712}{22370298575625}\,\*\z3
+\frac{615424}{9555}\,\*\z5\Big)
\nn \\[0.5mm] & & \mbox{\hspp}
+\nfs\,\*\cfs\,\*\Big(\frac{649724993957136005232607141}{339884061189952195917000000}
+\frac{178574416}{42567525}\,\*\z3
-\frac{13568}{4095}\,\*\z4\Big)
\nn \\[0.5mm] & & \mbox{\hspp}
+\nfs\,\*\ca\,\*\cf\,\*\Big(-\frac{2493111180935243429476603}{1493995873362427234800000}
-\frac{83451856}{36891855}\,\*\z3
+\frac{13568}{4095}\,\*\z4\Big)
\nn \\[0.5mm] & & \mbox{\hspp}
+\nft\,\*\cf\,\*\Big(\frac{321519364624645823}{538959550275045900}
-\frac{13568}{36855}\,\*\z3\Big)
\; , \label{Ggq3N14}
\\[2mm]
&& \hspn \gamma_{\,\rm gq}^{\,(3)}(N=16) \,=\,
\cff\,\*\Big(-\frac{91333794681488605852642048633486608590877607351}{431323535597978497478425306351127101440000000}
\nn \\[0.5mm] & & \mbox{\hspp}
-\frac{15461805996412648565848919}{58626866062381256640000}\,\*\z3
+\frac{105621100681}{1715313600}\,\*\z4
+\frac{2466482078}{5579145}\,\*\z5\Big)
\nn \\[0.5mm] & & \mbox{\hspp}
+\ca\,\*\cft\,\*\Big(\frac{69709103720766489723595997659712410526245706881}{161746325849241936554409489881672663040000000}
\nn \\[0.5mm] & & \mbox{\hspp}
+\frac{561806270069913957735019}{1503252975958493760000}\,\*\z3
-\frac{181207395103216571}{2126662954416000}\,\*\z4
-\frac{2706171203}{3550365}\,\*\z5\Big)
\nn \\[0.5mm] & & \mbox{\hspp}
+\cas\,\*\cfs\,\*\Big(-\frac{2905887663991964454299544802600912038519944459}{13478860487436828046200790823472721920000000}
\nn \\[0.5mm] & & \mbox{\hspp}
-\frac{14062285015933017481879}{142125735908803046400}\,\*\z3
+\frac{14216141064343763}{708887651472000}\,\*\z4
+\frac{93724168}{255255}\,\*\z5\Big)
\nn \\[0.5mm] & & \mbox{\hspp}
+\cat\,\*\cf\,\*\Big(-\frac{317835545673384782764570229718975673012928297}{53915441949747312184803163293890887680000000}
\nn \\[0.5mm] & & \mbox{\hspp}
-\frac{1262927249130586239359921}{39084577374920837760000}\,\*\z3
+\frac{475554692179667}{132916434651000}\,\*\z4
-\frac{23407688153}{937296360}\,\*\z5\Big)
\nn \\[0.5mm] & & \mbox{\hspp}
+\dfRAnr\,\*\Big(-\frac{11773700142311179915702121}{240974431125652224000000}
-\frac{101169081992668980737}{563503134009816000}\,\*\z3
+\frac{4137449651}{15621606}\,\*\z5\Big)
\nn \\[0.5mm] & & \mbox{\hspp}
+\nf\,\*\cft\,\*\Big(-\frac{57298339123449136718474401962230424513361}{7181544049250391233407014757761024000000}
-\frac{87782813700795439}{5546240318619000}\,\*\z3
\nn \\[0.5mm] & & \mbox{\hspp}
-\frac{16614041449}{1562160600}\,\*\z4
+\frac{1096}{51}\,\*\z5\Big)
+\nf\,\*\ca\,\*\cfs\,\*\Big(\frac{93344344164963857590569964407611700892711}{10772316073875586850110522136641536000000}
\nn \\[0.5mm] & & \mbox{\hspp}
-\frac{2475119092628220127}{88739845097904000}\,\*\z3
+\frac{643872057559}{23605982400}\,\*\z4
-\frac{548}{153}\,\*\z5\Big)
\nn \\[0.5mm] & & \mbox{\hspp}
+\nf\,\*\cas\,\*\cf\,\*\Big(-\frac{2033308368213027173938918062479854873193}{633665651404446285300618949214208000000}
\nn \\[0.5mm] & & \mbox{\hspp}
+\frac{15529936726131895757}{354959380391616000}\,\*\z3
-\frac{3535338880967}{212453841600}\,\*\z4
-\frac{121513}{7803}\,\*\z5\Big)
\nn \\[0.5mm] & & \mbox{\hspp}
+\nf\,\*\dfRRnr\,\*\Big(-\frac{405223386945234740808587}{9940195283933154240000}
-\frac{14956670451523}{1563722760600}\,\*\z3
+\frac{145816}{2601}\,\*\z5\Big)
\nn \\[0.5mm] & & \mbox{\hspp}
+\nfs\,\*\cfs\,\*\Big(\frac{9044520444067665064750987439226131}{5275271823213838538966191718400000}
+\frac{4119326989}{1171620450}\,\*\z3
-\frac{2192}{765}\,\*\z4\Big)
\nn \\[0.5mm] & & \mbox{\hspp}
+\nfs\,\*\ca\,\*\cf\,\*\Big(-\frac{60504685061989103630936679672439}{38788763405984106904163174400000}
-\frac{3125731858}{1757430675}\,\*\z3
+\frac{2192}{765}\,\*\z4\Big)
\nn \\[0.5mm] & & \mbox{\hspp}
+\nft\,\*\cf\,\*\Big(\frac{34108337373135323374301}{63317015347371757171200}
-\frac{2192}{6885}\,\*\z3\Big)
\; , \label{Ggq3N16}
\\[2mm]
&& \hspn \gamma_{\,\rm gq}^{\,(3)}(N=18) \,=\,
\cff\,\*\Big(-\frac{199441346676346613221100951050417553229794351172058063}{903627934773265885372610492741460297127036416000000}
\nn \\[0.5mm] & & \mbox{\hspp}
-\frac{9734182701413671124997104021}{35813961619291817562100500}\,\*\z3
+\frac{55191569546}{916620705}\,\*\z4
+\frac{555009226736}{1208442807}\,\*\z5\Big)
\nn \\[0.5mm] & & \mbox{\hspp}
+\ca\,\*\cft\,\*\Big(\frac{1703079039178221518412667538347685102671198180332127}{3752088310477228589782465713805371475960012800000}
\nn \\[0.5mm] & & \mbox{\hspp}
+\frac{58151956931435413049516329}{144995796029521528591500}\,\*\z3
-\frac{208211747676885206}{2413254243364965}\,\*\z4
-\frac{618848170736}{769009059}\,\*\z5\Big)
\nn \\[0.5mm] & & \mbox{\hspp}
+\cas\,\*\cfs\,\*\Big(-\frac{123319626596597732563913557669030322668202344255007}{521123376455170637469786904695190482772224000000}
\nn \\[0.5mm] & & \mbox{\hspp}
-\frac{97421341297611056710882606}{813953673165723126411375}\,\*\z3
+\frac{7737412238766461}{321767232448662}\,\*\z4
+\frac{1104257588788}{2819699883}\,\*\z5\Big)
\nn \\[0.5mm] & & \mbox{\hspp}
+\cat\,\*\cf\,\*\Big(\frac{115832786415748476638280610626833293796768491507}{55177769271723949849742142850078992293529600000}
\nn \\[0.5mm] & & \mbox{\hspp}
-\frac{395558870241334146068201903}{13482903197851037199849600}\,\*\z3
+\frac{9748563515611381}{4826508486729930}\,\*\z4
-\frac{214104770108}{8459099649}\,\*\z5\Big)
\nn \\[0.5mm] & & \mbox{\hspp}
+\dfRAnr\,\*\Big(-\frac{1534254855736463458285860487597}{32941608367048794488275200000}
-\frac{263063336077755987407089}{1556198430038208356400}\,\*\z3
\nn \\[0.5mm] & & \mbox{\hspp}
+\frac{705610821248}{2819699883}\,\*\z5\Big)
+\nf\,\*\cft\,\*\Big(-\frac{154871348508574160288273783965897373976839491}{31860889749307135877273684882463194351308800}
\nn \\[0.5mm] & & \mbox{\hspp}
-\frac{18100451697075180817}{1062928800827568675}\,\*\z3
-\frac{84572797384}{8459099649}\,\*\z4
+\frac{55040}{2907}\,\*\z5\Big)
\nn \\[0.5mm] & & \mbox{\hspp}
+\nf\,\*\ca\,\*\cfs\,\*\Big(\frac{494213149462187941188004051358795571999087887}{93708499262668046697863779066068218680320000}
\nn \\[0.5mm] & & \mbox{\hspp}
-\frac{197585480953102482619}{9566359207448118075}\,\*\z3
+\frac{59943194135296}{2410843399965}\,\*\z4
-\frac{27520}{8721}\,\*\z5\Big)
\nn \\[0.5mm] & & \mbox{\hspp}
+\nf\,\*\cas\,\*\cf\,\*\Big(-\frac{936013096859225678199723846435281783145737}{290119192763678163151281049740149283840000}
\nn \\[0.5mm] & & \mbox{\hspp}
+\frac{122057786649260319779}{3237844654828593810}\,\*\z3
-\frac{35839946880856}{2410843399965}\,\*\z4
-\frac{20446720}{1491291}\,\*\z5\Big)
\nn \\[0.5mm] & & \mbox{\hspp}
+\nf\,\*\dfRRnr\,\*\Big(-\frac{2376495270782792811304983137}{65883216734097588976550400}
-\frac{9726278907372152}{1143120431067615}\,\*\z3
+\frac{24663040}{497097}\,\*\z5\Big)
\nn \\[0.5mm] & & \mbox{\hspp}
+\nfs\,\*\cfs\,\*\Big(\frac{2076988113916115506387549621182589219}{1350630257073663390768162168309360000}
+\frac{13478091680}{4478346873}\,\*\z3
-\frac{22016}{8721}\,\*\z4\Big)
\nn \\[0.5mm] & & \mbox{\hspp}
+\nfs\,\*\ca\,\*\cf\,\*\Big(-\frac{51603554118665492025456278063355347}{35310594956174206294592474988480000}
-\frac{543557988224}{380659484205}\,\*\z3
+\frac{22016}{8721}\,\*\z4\Big)
\nn \\[0.5mm] & & \mbox{\hspp}
+\nft\,\*\cf\,\*\Big(\frac{25322735830563737715837587}{51572104731780217289424720}
-\frac{22016}{78489}\,\*\z3\Big)
\; , \label{Ggq3N18}
\\[2mm]
&& \hspn \gamma_{\,\rm gq}^{\,(3)}(N=20) \,=\,
\cff\,\*\Big(-\frac{18011842890292937504270512411100571563413211395232599}{78747532442114674106545576709495450730024960000000}
\nn \\[0.5mm] & & \mbox{\hspp}
-\frac{27623613597125033101695488051}{98312835817663812915570000}\,\*\z3
+\frac{57192203098}{972173475}\,\*\z4
+\frac{923526447854}{1935088155}\,\*\z5\Big)
\nn \\[0.5mm] & & \mbox{\hspp}
+\ca\,\*\cft\,\*\Big(\frac{742257701799954265313171119485669995895179080771700597}{1566661434901018253277590947167856861892075520000000}
\nn \\[0.5mm] & & \mbox{\hspp}
+\frac{304638383107094264933207}{710296405760118870000}\,\*\z3
-\frac{570786749229593761}{6585879026727000}\,\*\z4
-\frac{149116082954}{175917105}\,\*\z5\Big)
\nn \\[0.5mm] & & \mbox{\hspp}
+\cas\,\*\cfs\,\*\Big(-\frac{13595171979898487398112366060650503619924613365949}{53542769477136645703266949663973235197952000000}
\nn \\[0.5mm] & & \mbox{\hspp}
-\frac{13706603212061214884924310847}{98312835817663812915570000}\,\*\z3
+\frac{1068938712970319969}{39515274160362000}\,\*\z4
+\frac{178657350101}{430019590}\,\*\z5\Big)
\nn \\[0.5mm] & & \mbox{\hspp}
+\cat\,\*\cf\,\*\Big(\frac{8128052126051322869747930125993834197558195023}{996276944019023178954993747070209054720000000}
\nn \\[0.5mm] & & \mbox{\hspp}
-\frac{4468686144075569343986511569}{165579512956065369120960000}\,\*\z3
+\frac{494114186999099}{627226573974000}\,\*\z4
-\frac{10876666809}{430019590}\,\*\z5\Big)
\nn \\[0.5mm] & & \mbox{\hspp}
+\dfRAnr\,\*\Big(-\frac{9934247787263135793881744121471193}{223044635561617655771521536000000}
-\frac{20055818840172257506705213}{125439024966716188728000}\,\*\z3
\nn \\[0.5mm] & & \mbox{\hspp}
+\frac{153050440696}{645029385}\,\*\z5\Big)
+\nf\,\*\cft\,\*\Big(-\frac{941248780576101712647404838543624320767081907}{394884456817228610955910317701869858560000000}
\nn \\[0.5mm] & & \mbox{\hspp}
-\frac{6339794916091394356451}{354748373774649855000}\,\*\z3
-\frac{91361957041}{9675440775}\,\*\z4
+\frac{6752}{399}\,\*\z5\Big)
\nn \\[0.5mm] & & \mbox{\hspp}
+\nf\,\*\ca\,\*\cfs\,\*\Big(\frac{102664753783997698300409838737001288852130553}{40379162501611346684213385870266391552000000}
\nn \\[0.5mm] & & \mbox{\hspp}
-\frac{5347393544742883698797}{354748373774649855000}\,\*\z3
+\frac{2584501408063}{112880142375}\,\*\z4
-\frac{3376}{1197}\,\*\z5\Big)
\nn \\[0.5mm] & & \mbox{\hspp}
+\nf\,\*\cas\,\*\cf\,\*\Big(-\frac{465634445017831591325444801806925052804292331}{148765335532252329889207211100981442560000000}
\nn \\[0.5mm] & & \mbox{\hspp}
+\frac{3171458341939781449}{96255155006010000}\,\*\z3
-\frac{4555835727754}{338640427125}\,\*\z4
-\frac{1539424}{125685}\,\*\z5\Big)
\nn \\[0.5mm] & & \mbox{\hspp}
+\nf\,\*\dfRRnr\,\*\Big(-\frac{2704920902087866223629472523102433}{83641738335606620914320576000000}
-\frac{108318373596246439}{14144672000584125}\,\*\z3
\nn \\[0.5mm] & & \mbox{\hspp}
+\frac{1863808}{41895}\,\*\z5\Big)
+\nfs\,\*\cfs\,\*\Big(\frac{4773708126683882394527205560488250927}{3460438305705072216870062548871040000}
+\frac{4000788628}{1527701175}\,\*\z3
\nn \\[0.5mm] & & \mbox{\hspp}
-\frac{13504}{5985}\,\*\z4\Big)
+\nfs\,\*\ca\,\*\cf\,\*\Big(-\frac{11325534673549584438954822499666967}{8259788293841919601074263155200000}
-\frac{11266215274}{9675440775}\,\*\z3
\nn \\[0.5mm] & & \mbox{\hspp}
+\frac{13504}{5985}\,\*\z4\Big)
+\nft\,\*\cf\,\*\Big(\frac{23642031766087097685137941}{52433512436087366888304000}
-\frac{13504}{53865}\,\*\z3\Big)
\; . \label{Ggq3N20}
\eea

}}
{\sc Form} files with the results for $\gamma_{\,\rm gq}^{}(N)$ at
even $N \leq 20$ and the partial all-$N$ expressions in the main text
have been deposited at the preprint server {\tt http://arXiv.org}
together with a {\sc Fortran} subroutine of our approximations for the 
splitting function $P_{\,\rm gq}^{\,(3)}(x)$ in 
eqs.~(\ref{eq:Pgq30-nf}) - (\ref{eq:Pgq3A3-nf5}).
These files are also available from the authors upon request.

\end{document}